\documentclass[preprintnumbers,amsmath,amssymb]{revtex4}

\usepackage{graphicx}
\usepackage{bm}

\DeclareMathOperator{\sign}{sign}

\begin{document}

\title{Localised time-periodic solutions of discrete nonlinear Klein-Gordon systems with convex on-site potentials}

\author{Dirk Hennig}
%{Dirk Hennig\\E-mail:hennigd@physik.hu-berlin.de}
%\affiliation{Humboldt Universit\"at zu Berlin, Institut f\"ur Physik, Germany} 
%Practice for Internal Medicine, Karl-Liebknecht-Stra{\ss}e 3, Eberswalde, Germany}

%\href{mailto:name.surname@gmail.com}{name.surname@gmail.com}

\date{\today}

\begin{abstract}
\noindent {\cal {\bf Abstract:}} The existence of nonzero  localised periodic solutions for 
 general one-dimensional discrete nonlinear Klein-Gordon systems 
 with convex on-site potentials is proved.  
The existence problem of localised solutions 
is expressed in terms of a fixed point equation for an operator on some appropriate function space  
 which is solved by means of Schauder's Fixed Point Theorem. 
\end{abstract}

%\pacs{05.45.-a, 63.20.Pw, 45.05.+x, 63.20.Ry}
\maketitle

\noindent 

Non-uniform structures  in nonlinear discrete lattice systems in the form of spatially
localised and time-periodically varying states, referred to as breather solutions, have attracted considerable interest. 
Rigorous results regarding   proofs of existence and stability of intrinsic localised time-periodic modes (also known as breathers) were 
provided in \cite{MacKay}-\cite{b12}. Several methods and techniques were used such as continuation of (trivially localised) solutions from the anti-integrable 
of zero couplings for sufficiently weak couplings, center manifold reduction methods,   Lyapunov-Schmidt decomposition and continuum approximation techniques, variational approaches, perturbative arguments for the characteristic exponents of the Floquet monodromy matrices and normal form theorems and Nekoroshev-type proofs. 
With this work we present a comparatively concise existence proof for breathers 
in general discrete nonlinear Klein-Gordon systems with convex on-site potential.

We discuss  the existence  of  localised time-periodic solutions of  
 general discrete nonlinear Klein-Gordon systems on 
infinite  lattices given by the following set of coupled oscillator equations
\begin{equation}
\frac{d^2 q_n}{dt^2}=\kappa\,[q_{n+1}-2q_n+q_{n-1}]-
V^{\prime}(q_n),\,\,\, n\in {\mathbb{Z}},\label{eq:system}
\end{equation}
where the $q_n(t)$ describe the coordinate of the $n$th oscillator at time $t$ and 
the prime $^{\prime}$ in the last term on the r.h.s. of (\ref{eq:system}) stands for the derivative with respect to $q_n$.
Each oscillator interacts with its neighbours to the left and right and the
strength of the  interaction 
is determined by the value of the parameter $\kappa$. 
This system has a Hamiltonian structure related to the energy
\begin{equation}
 H=\sum_{n\in {\mathbb{Z}}}\left(\frac{1}{2}p_n^2+V(q_n)+
 \frac{\kappa}{2}(q_{n+1}-q_{n})^2\right),\label{eq:Ham}
\end{equation}
 and it is time-reversible with respect to the involution $p\mapsto -p$.

\vspace*{0.5cm}

Assume that the following condition on the on-site potential $V(x)$ holds:

\vspace*{0.5cm}

\noindent {\bf{A:}}  
$V:\,\,{\mathbb{R}}\,\rightarrow\,{\mathbb{R}}$ is even, $V(x)=V(-x)$, 
non-negative and at least twice continuously 
differentiable and uniformly convex which is characterised by the following properties:
There exists two constants $0<\underline{K}<\overline{K}<\infty$ 
such that
\begin{equation}
 \underline{K}\le V^{\prime \prime}(x)\le \overline{K},\,\,\, \forall x 
 \in {\mathbb{R}}.\label{eq:assumptions1}
\end{equation}
Then, one has
\begin{equation}
 \frac{1}{2}\underline{K}\,x^2\le V(x) \le \frac{1}{2}\overline{K}\,x^2,\,\,\,
 |V^{\prime}(x)|\le \overline{K}|x|,\,\,\, \forall x \in {\mathbb{R}}.
\end{equation}
Further assume that 
\begin{equation}
V(0)=V^{\prime}(0)=0,\,\,\,V^{\prime\prime}(0)=1.\label{eq:assumptions}
\end{equation}

That is, the unique equilibrium at $x=0$ is a global minimum of $V(x)$ and $V^{\prime}(x)$ can be expressed as
\begin{equation}
 V^{\prime}(x)=x+W^{\prime}(x),
\end{equation}
with non-negative $W(x)$ and $\sign(V^{\prime}(x))=\sign(W^{\prime}(x))$.
Hence, 
\begin{equation}
 V(x)=\frac{1}{2}x^2+W(x).\label{eq:Af}
\end{equation}
\vspace*{0.5cm}

The solutions of the system obtained by linearising  equations (\ref{eq:system}) around the equilibrium $q_n=0$ are superpositions of plane wave solutions (phonons) 
\begin{equation}
 q_n(t)= \exp(i(kn-\omega t)),
\end{equation}
with frequencies 
\begin{equation}
 \omega^2(k) =1+4\kappa\sin^2\left(\frac{k}{2}\right),\,\,\,k\in[-\pi,\pi].
\end{equation}
These (extended) states disperse. Therefore the  frequency $\Omega$ of a  localised time-periodic solution must satisfy the non-resonance condition $\Omega \neq |\omega(k)|/m$ for any integer $m\ge 1$. This requires 
$\Omega^2 >1+4\kappa$ as a necessary condition for the existence of localised time-periodic solutions of system (\ref{eq:system}).

In the following we prove 
the existence  of  localised periodic 
solutions of system (\ref{eq:system}) on infinite lattices.

To this end, some appropriate function spaces are introduced on which the original 
problem is presented as a fixed point problem for a corresponding operator. 
Utilising Schauder's 
Fixed Point Theorem we establish the existence of localised solutions in a similar vein to the approach in \cite{SIAM} 
%Such an approach was used in \cite{Mirollo} 
treating the existence of travelling wave solutions in finite arrays of coupled Josephson junction systems.

\vspace*{0.5cm}

With concern to the existence of localised periodic solutions to system (\ref{eq:system}) we state:

\vspace*{0.5cm}
\noindent{\bf Theorem:}{\it \,\,Let {\bf A}  hold.  Then 
 there exist nonzero sequences  $q\equiv \{q_n\}_{n\in {\mathbb{Z}}} 
 \in H^2([0,T];l^2)$ with zero time average $\int_0^T q_n(t)dt=0,$ that are  localised time-periodic solutions of system (\ref{eq:system}) 
with period   $T=\frac{2\pi}{\Omega}$ 
provided
\begin{equation}
\Omega^2> 1+4\kappa\label{eq:As}.
\end{equation}
}

\vspace*{0.75cm}

For the  proof of the assertions of the Theorem  the strong version of Schauder's Fixed Point Theorem is used \cite{Schauder}: 
{\it Let $G$ be a closed convex subset of a
Banach space and $f$ a  continuous map of $G$ into a compact subset of $G$. Then  $f$ has a fixed point.}

\vspace*{1.0cm}

\noindent{\bf Proof:} For the following considerations Eq.\,(\ref{eq:system}) is suitably re-written as:
\begin{equation}
 \ddot{q}_n+q_n- \kappa [q_{n+1}-2q_n+q_{n-1}]=-W^{\prime}(q_n)
\label{eq:ref1}
\end{equation}
Thus only the r.h.s. of  (\ref{eq:ref1}) features terms nonlinear in $q$. Ultimately, 
(\ref{eq:ref1}) shall be re-formulated as a fixed point equation in $q$. 

We introduce appropriate function spaces.
Denote by 
\begin{equation}
 {X}^{0} =  \left\{ q\in L^{2}_{per}((0,T);{l}^2)\,\vert\,\int_0^T q_n(t)dt=0,\,\,\, n\in {\mathbb{Z}}
 \right\}
\end{equation}
 the space of $T-$periodic square integrable
 functions of time with zero time average assuming values in $l^2$. $X^0$ is a closed convex subset of $L^{2}_{per}((0,T);{l}^2)$. We introduce further the Sobolev space $X^2=\{q\in H^{2}_{per}((0,T);{l}^2)\}$
 containing  $T-$periodic 
  functions of time  assuming values in $l^2$ which, 
 together with their weak derivative $D^{\alpha\le2}$ up to order 
$2$, are in $X^0$. The following norms are used:
\begin{eqnarray}
\parallel q \parallel_{X^0}^2&=&\frac{1}{T}\int_{0}^{T}\,\parallel q(t)\parallel^2_{{l}^2}dt=
\sum_{m\in {\mathbb{Z}}\setminus \{0\}} \frac{1}{2\pi} \int_{0}^{2\pi}\left|\hat{\tilde{q}}_{k,m}\right|^2 dk\\
 \parallel q \parallel_{X^2}^2&=&\frac{1}{T}\int_{0}^{T}\,\left(\parallel q(t)\parallel^2_{{l}^2}
 +\parallel D{q}(t)\parallel^2_{{l}^2}+\parallel D^2{q}(t)\parallel^2_{{l}^2}\right)dt=
 \sum_{m\in {\mathbb{Z}}\setminus \{0\}}\left(1+(m\Omega)^2+(m\Omega)^4\right)
 \frac{1}{2\pi} \int_{0}^{2\pi}
 |\hat{\tilde{q}}_{m,k}|^2dk,\label{eq:norml2}
\end{eqnarray} 
where $\hat{q}_{n,k}$ determines the temporal Fourier-transform:
\begin{eqnarray}
 q_n(t)&=&\sum_{m\in {\mathbb{Z}}\setminus \{ 0 \}}
\hat{q}_{n,m} \exp\left( i\Omega m t\right),\,t\in [0,T],\,\,\,\,\,\,{\hat{q}}_{n,m}=\frac{1}{T} \int_{0}^{T} q_n(t)\exp(-i \Omega m t)dt,\\
\hat{q}_{n,m}&=&\overline{\hat{q}}_{n,-m},
\label{eq:FTX2}
\end{eqnarray}
and $\tilde{q}_k$ is the spatial Fourier-transform:
 \begin{eqnarray}
 q_n(t)&=&\frac{1}{2\pi}\int_{0}^{2\pi}\tilde{q}_k(t)\exp(ikn),\,\,\,\,\,\tilde{q}_k(t)=\sum_{n}q_n(t)\exp(-i\,k\,n),\label{eq:FTspatial}\\
 \tilde{q}_{k}&=&\overline{\tilde{q}}_{-k},
\end{eqnarray}
of $q_n(t)$, respectively, and $\overline{u}$ denotes the conjugate of 
$u$.

The Hilbert space of square-summable sequences, ${l}^2(\mathbb{Z})$, is endowed with the standard scalar product:
\begin{equation}
 \left<q,p\right>:=\sum_{n}\,q_n\,p_n\,,\qquad \parallel q \parallel^2_{l^2}:=\left<q,q\right>\,.\label{eq:norm}
\end{equation}

For later use we note that $X^2$ is compactly embedded in $X^0$ ($X^2 \Subset X^{0}$).

We relate the l.h.s. of (\ref{eq:ref1}) to  the 
linear mapping: $M\,:\,X^2\,\rightarrow\,X^{0}$:
\begin{equation}
 M(u_n)=\ddot{u}_n+{u}_n- \kappa [u_{n+1}-2u_n+u_{n-1}].  
\end{equation}
Using the relations for the temporal and spatial Fourier-transforms (\ref{eq:FTX2}),(\ref{eq:FTspatial}), 
one obtains the representation
\begin{equation}
 u_n(t)=\sum_{m\in {\mathbb{Z}}\setminus \{0\}}\frac{1}{2\pi} \int_{0}^{2\pi} {\hat{\tilde{u}}}_{k,m}\exp(ikn)dk\exp(i\Omega mt).\label{eq:FTst}
\end{equation}

Applying the operator $M$ to the Fourier elements $\exp(ikn)\exp(i\,\Omega m t)$ 
in (\ref{eq:FTst}), results in
\begin{equation}
 M\exp(ikn)\exp(i\,\Omega mt)=\nu_m(k) \exp(ikn)\exp(i\,\Omega mt),
\end{equation}
where 
\begin{equation}
 \nu_m(k)= - \Omega^2\,m^2+1+4\,\kappa\sin^2\left(\frac{k}{2}\right).
\end{equation}

As the hypothesis (\ref{eq:As}) guarantees that $\nu_m(k) \ne 0$, $\forall \,\,m\in {\mathbb{Z}} $ and $\forall \,\,k\in [0,2\pi] $, the mapping $M$ possesses an inverse obeying 
$M^{-1}\exp(i\,\Omega lt)=(1/\nu_l)\exp(i\,\Omega lt)$ and for the norm of the  linear operator $M^{-1}:\,\,X^{0} \rightarrow X^2$ one derives using the hypothesis  (\ref{eq:As}) the upper bound:
\begin{eqnarray}
\parallel M^{-1} \parallel_{X^{0},X^2}&=&
 \sup_{0 \neq q \in X^0}\frac{\parallel M^{-1}\,q \parallel_{X^2}}
{\parallel q \parallel_{X^0}}\nonumber\\
 & =&\sup_{0 \neq q \in X^0}\frac{\left(\int_0^T\,\left[\parallel M^{-1} q(t)\parallel^2_{{l}^2}
 +\parallel D{M^{-1}q(t)} 
 \parallel^2_{{l}^2}+\parallel 
 D^2{M^{-1}q(t)}
 \parallel^2_{{l}^2}\right]dt \right)^{1/2}}
 {\parallel q \parallel_{X^0}}\nonumber\\
&=&\sup_{0 \neq q \in X^0}
\frac{\left(
\sum_{m^\prime}\frac{1}{2\pi} \int_{0}^{2\pi}\left|\frac{\hat{\tilde{q}}_{k,m}}{\nu_m(k)}\right|^2 dk
+\sum_{m^\prime}\frac{1}{2\pi} \int_{0}^{2\pi}\left|\frac{i \Omega m \hat{\tilde{q}}_{k,m}}{\nu_m(k)}\right|^2 dk
+\sum_{m^\prime}\frac{1}{2\pi} \int_{0}^{2\pi}\left|\frac{(i \Omega m)^2 \hat{\tilde{q}}_{k,m}}{\nu_m(k)}\right|^2 dk
\right)^{1/2}}{\parallel q \parallel_{X^0}}\nonumber\\
 &\le&\sup_{m\in {\mathbb{Z}}\setminus \{ 0 \}}\sup_{k\in [0,2\pi]}{\frac{1}{|\nu_m(k)|}} 
 \cdot\frac{\left(\sum_{m^\prime} \left(1+(\Omega\,m)^2+(\Omega\,m)^4\right) \frac{1}{2\pi} \int_{0}^{2\pi}
 \left|\hat{\tilde{q}}_{k,m}\right|^2 dk\right)^{1/2}}{\parallel q \parallel_{X^0}}\nonumber\\
&\le&  \frac{1}
 {\Omega^2-(1+4\,\kappa)}< \infty,\label{eq:boundL}
\end{eqnarray}
verifying boundedness of $M^{-1}$,  
and we used the notation $\sum_{l^\prime} = \sum_{l\in {\mathbb{Z}}\setminus \{ 0 \}}$.

Furthermore associated with the r.h.s. of (\ref{eq:ref1}) we introduce 
the nonlinear operator $N\,:\,X^{0}\,\rightarrow\,X^{0}$, as
\begin{equation}
 (N(u))_n= -W^{\prime}(u_n).\label{eq:G}
\end{equation}
   
In order to show  that the  operator $N$ 
is continuous on $X^{0}$ we prove that  
 $N$ is Frechet differentiable at any $u$ with bounded derivative. In detail we have
\begin{equation}
 N^{\prime}(u):\,h\in X^{0}\,\mapsto N^{\prime}(u)[h] =-W^{\prime\prime}(u)h \in X^{0},
\end{equation}
so that
\begin{eqnarray}
 \parallel N^{\prime}(u)[h]\parallel_{X^{0}}^2&=&
 \int_0^T
 \sum_{n \in \mathbb{Z}} \left|\,-W^{\prime\prime}(u_n(t))\,h_n(t)\right|^2\,dt\nonumber\\
  &\le& 
  \int_0^T
 \sum_{n \in \mathbb{Z}} |1-\,V^{\prime\prime}(u_n(t))|^2\,|h_n(t)|^2\,dt\nonumber\\
&\le& (1+\overline{K}^2)\,\parallel h\parallel^2_{X^0},
\end{eqnarray}
proving the (uniform) boundedness of the differential. Hence, 
\begin{equation}
 \parallel N^{\prime}(u)\parallel_{{\cal{L}}(X^{0},X^{0})}\le \sqrt{1+\overline{K}^2},\label{eq:Frechet}
\end{equation}
viz. the derivative is bounded and for  the range of $N$ in $X^0$ we get the bound
\begin{eqnarray}
 \parallel N(u) \parallel_{X^{0}}^2&=&\int_0^T
 \sum_{n \in \mathbb{Z}} |\,-W^{\prime}(u_n(t))|^2\,dt\nonumber\\
 &\le& 
 \int_0^T
 \sum_{n \in \mathbb{Z}} |\,u_n(t)-V^{\prime}(u_n(t))|^2\,dt\nonumber\\
 &\le&
 (1+\overline{K}^2)\int_0^T\sum_{n \in \mathbb{Z}}|u_n(t)|^2\,dt\le  (1+\overline{K}^2) \parallel u\parallel^2_{X^0}.
 \label{eq:rangeN}
\end{eqnarray}

Finally, we express the problem  (\ref{eq:ref1})
 as a fixed point equation in terms of a mapping $X^0\,\rightarrow\, X^{2} \Subset X^{0}$:
\begin{equation}
 q=M^{-1}\circ N(q)\equiv S(q).\label{eq:compose}
\end{equation}

We verify that the range of $S$ is contained in a compact subset of $X^0$. 
Representing  $N(q) \in X^0$ in terms of its spatial and temporal Fourier-transforms  as
%given by  $p=(p)_{n\in \mathbb{Z}}$ where
\begin{equation}
 (N(q))_n(t)=\sum_{m\in {\mathbb{Z}}\setminus \{ 0 \}} \frac{1}{2\pi} \int_{0}^{2\pi}
\hat{\tilde{N}}_{k,m}\exp(ikn)dk \exp\left(i\Omega m t\right),\label{eq:FT}
\end{equation}
 the Fourier coefficients of $M^{-1}(N(q))$ fulfill $\forall q\in X^0$, $\forall k\in[0,2\pi],\,\,\,\forall m\in {\mathbb{Z}}\setminus \{ 0 \}$
\begin{equation}
 \left|\frac{\hat{{\tilde{N}}}_{k,m}}{- \Omega^2\,m^2+1+4\,\kappa\sin^2\left(\frac{k}{2}\right)}\right|^2\le 
 \frac{\sum_{m\in {\mathbb{Z}}\setminus \{ 0 \}}
 \frac{1}{2\pi} \int_{0}^{2\pi}\left|\hat{{\tilde{N}}}_{k,m}\right|^2dk}{m^4[\Omega^2-(1+4\,\kappa)]^2}=\frac{\parallel N(q) \parallel_{X^{0}}^2}{m^4[\Omega^2-(1+4\,\kappa)]^2}.\label{eq:FcN}
\end{equation}

Exploiting the coercivity  of the conserved total energy $E=H$ in (\ref{eq:Ham}) and the expression (\ref{eq:Af})  one can 
  bound the $l^2$-norm of $q=\{q_n\}_{n\in {\mathbb{Z}}}$ as follows $\parallel q(t)\parallel_{l^2}^2 \le 2E$ for every $t\in [0,T]$, so that  
  \begin{equation}
 \parallel {q}\parallel_{X^0}^2\le 2E,
 \end{equation}
 from which, using (\ref{eq:rangeN}), we deduce
 \begin{equation}
  \parallel N(q) \parallel_{X^{0}}^2\le 2(1+\overline{K}^2)E. \label{eq:NE}
 \end{equation}

Hence, from (\ref{eq:FcN}) and (\ref{eq:NE}) one concludes that $S$ maps $X^0$ into the subset 
\begin{equation}
 Y^0=\left\{p=\left\{p\right\}_{n\in {\mathbb{Z}}} \in X^0,\,p_n(t)=\sum_{m\in {\mathbb{Z}}\setminus \{ 0 \}}
 \frac{1}{2\pi} \int_{0}^{2\pi}\hat{{\tilde{p}}}_{k,m}\,\exp(ikn)dk\exp\left( i \Omega m t\right)\,\left|\,| \hat{{\tilde{p}}}_{k,m}|\le \frac{M}{m^2}\right.\right\},
\end{equation}\label{eq:range}
which is, for every
\begin{equation}
M=\frac{\sqrt{2(1+\overline{K}^2)E}}{\Omega^2-(1+4\,\kappa)}>0 
\end{equation}
compact in $X^0$.

That is, the operator $S$ maps closed convex subsets $X^0$ of $L_{per}^2((0,T);l^2)$ into compact subsets $Y^0$ of $X^0$.

\vspace*{0.5cm}

Clearly $S$ is continuous on $X^{0}$ as the relations (\ref{eq:boundL}) and (\ref{eq:Frechet}) establish that its constituents $M^{-1}$ and $N$ are continuous.
 Schauder’s fixed point theorem
implies then that the fixed point equation $q =  S(q)$ has at least one solution.
Moreover,  by the hypothesis (\ref{eq:As}) the values of the 
 frequency of oscillations $\Omega$ lie above the upper edge  
of the continuous (phonon) spectrum  determined by $1+4\kappa$. Thus the corresponding solutions 
have to be anharmonic which necessitates nonzero nonlinearity. 
Only the latter facilitates amplitude-depending  tuning of the frequency of oscillations so that the breather frequency $\Omega$
lies outside of the phonon spectrum.
Thus it must  hold that  
 $\parallel N(q)\parallel_{Y^0} =\parallel -W^{\prime}(q)\parallel_{Y^0}\not \equiv 0$, which is by {\bf{A}} satisfied if and only if  
 $q \not \equiv 0$.
 That is, the fixed point equation (\ref{eq:compose}) possesses a  nonzero 
solution   and the proof is finished.

\vspace*{0.5cm}

\hspace{16.5cm} $\square$

\vspace*{1.0cm}

{\it Remarks:}  (i) As the obtained time-periodic $H^2$ fixed-point-solutions are by Sobolev embeddings $C^1$ in time and since the operator $q \mapsto  W^{\prime}(q)$ maps $C^1$ into itself, one concludes from Eq.\,(\ref{eq:ref1}) that $\ddot{q} \in C^1$  so that the solutions are in fact $C^3$ in time and hence, are classical solutions. 

(ii) The localised solutions on the infinite lattice ${\mathbb{Z}}$ are represented 
by (infinite) square-summable sequences, viz. decay of the states for $|n|\rightarrow \infty$ takes place in the sense of the 
$l^2$ norm. This approach allows for general localised patterns (e.g. multi-site breathers). 
Regarding so-called single-site breathers, their exponential decay can be directly incorporated in our existence proof  by using an exponentially  weighted norm instead of the standard one in (\ref{eq:norm}) and the corresponding existence proof proceeds in the same manner as the general one above. (See also the relevant discussion in \cite{MacKay}.)

\vspace*{1.0cm}

In summary, we have proven the existence  of nonzero localised periodic  solutions 
for  general Klein-Gordon systems with convex on-site potentials on the lattice ${\mathbb{Z}}$.
The existence problem has been 
reformulated  
as a fixed point problem for an operator on a function space  
which is solved with the help  of Schauder's Fixed Point Theorem.  
% A critical energy threshold below which only  the existence of  trivial  localised periodic  solutions  is possible, has been obtained by means of  the Contraction Mapping Principle.
The method employed in this paper can be straightforwardly extended to treat also the general Klein-Gordon systems 
on lattices of higher dimension.

\end{document}